\documentclass[11pt,aps,prd,preprint,superscriptaddress]{revtex4}
\usepackage{amsmath}
\usepackage[dvips]{graphicx}
\usepackage[english]{babel}
\newcommand{\be}{\begin{equation}}
\newcommand{\ee}{\end{equation}}
\newcommand{\ben}{\begin{eqnarray}}
\newcommand{\een}{\end{eqnarray}}

\begin{document}
\title{Observational constraints on a holographic, interacting dark energy model}
\author{Iv\'{a}n Dur\'{a}n\footnote{E-mail: ivan.duran@uab.cat}}
\affiliation{Departamento de F\'{\i}sica, Universidad Aut\'{o}noma
de Barcelona, 08193 Bellaterra (Barcelona), Spain.}
\author{Diego Pav\'{o}n\footnote{E-mail: diego.pavon@uab.es}}
\affiliation{Departamento de F\'{\i}sica, Universitat Aut\`{o}noma
de Barcelona, 08193 Bellaterra (Barcelona), Spain.}
\author{Winfried Zimdahl\footnote{E-Mail: winfried.zimdahl@pq.cnpq.br}}
\affiliation{Departamento de F\'{\i}sica, Universidade Federal do
Espirito Santo, Brasil.}

\begin{abstract}
We constrain an interacting, holographic dark energy model, first
proposed by two of us in \cite{wd-cqg}, with observational data
from supernovae, CMB shift, baryon acoustic oscillations, x-rays,
and the Hubble rate. The growth function for this model is also
studied. The model fits the data reasonably well but still the
conventional $\Lambda$CDM model fares better. Nevertheless, the
holographic model greatly alleviates the coincidence problem and
shows compatibility at $1\sigma$ confidence level with the age of
the old quasar APM 08279+5255.
\end{abstract}
\maketitle

\section{Introduction}
Models of holographic dark energy have become popular as they rest
on the very reasonable assumption that the entropy of every
bounded region of the Universe, of size $L$, should not exceed the
entropy of a Schwarzschild black hole of the same size. That is to
say,
\be L^{3}\, \Lambda^{3} \leq S_{BH} \simeq L^{2}M_{Pl}^{2}\,
\quad\qquad (M_{Pl}^{2} = (8\pi G)^{-1})\, ,
 \label{bound1}
\ee
where $\Lambda$ stands for the ultraviolet cutoff, the infrared
cutoff is set by $L$.

However, as demonstrated by Cohen {\it et al.} \cite{cohen}, an
effective field theory that saturates the above inequality
necessarily includes states for which the Schwarzschild radius
exceeds $L$. It is therefore natural to replace the said bound by
another one that excludes such states right away, namely,
\be L^{3}\, \Lambda^{4} \leq M_{Pl}^{2}\, L \, .  \label{bound2}
\ee
This bound guarantees that the energy $L^{3}\Lambda^{4}$ in a
region of the size $L$ does not exceed the energy of a black hole
of the same size \cite{mli1}. By saturating the inequality
(\ref{bound2}) and identifying $\Lambda^{4}$ with the density of
holographic dark energy, $\rho_{X}$, it follows that
\cite{cohen,mli1}
\be \rho_{X}= \frac{3 c^{2}}{8\pi G \, L^{2}}\, , \label{rhox} \ee
where the factor $3$ was introduced for convenience and $c^{2}$ is
a dimensionless quantity, usually assumed constant, that collects
the uncertainties of the theory (such as the number particle
species and so on). For a more thorough motivation of holographic
dark energy see Section 3 of \cite{wd-cqg}.

Last relationship is widely used in setting models of holographic
dark energy that aim to explain the present stage of cosmic
accelerated expansion, \cite{Riess-1998,saul,komatsu,Amanullah},
via the huge negative pressure associated to them. Broadly
speaking holographic dark energy models fall into three main
groups depending on the choice of the infrared cutoff, $L$.
Namely, the Hubble radius \cite{dw-plb,wd-cqg}, the event horizon
radius \cite{mli1,wang-elcio,wang-lin,setare,mli2,micheletti}, and
the Ricci's length \cite{gao,suwa,xu,xu-wang}. The particle
horizon radius was also used \cite{Hsu} but it presents the severe
drawback of leading to a cosmology incompatible with a transition
from deceleration to acceleration during the Universe expansion.

In this paper we consider a spatially flat Friedmann-
Robertson-Walker universe dominated by holographic dark energy
(with the infrared cutoff set by the Hubble radius, i.e., $L =
H^{-1}$) and pressureless dark matter such that these two
components are dynamically linked by an interaction term. The
model was introduced in \cite{wd-cqg}. Here we constrain it with
data from supernovae type Ia (SN Ia), the shift of the first
acoustic peak in the cosmic background radiation (CMB shift),
baryon acoustic oscillations (BAO), and x-rays (strongly related
to the baryon gas abundance in galaxy clusters), and Hubble's
history, $H(z)$. We also study the evolution of the growth
function which potentially may constrain the model as well. But we
do not use them because, at present, these data are far noisier
than those in the other data sets.

The paper is organized as follows. Section II recalls the
holographic interacting model. Section III constrains the model
with observational data. Notwithstanding it does not contains the
flat $\Lambda$CDM model as a limiting case (at variance with,
e.g., quintessence models) it shows a sizable overlap with the
latter. Section IV studies the growth function. Finally, Section V
summarizes our overall conclusions. As usual, a zero subindex
indicates the present value of the corresponding quantity.


%
\section{Basics of the model}
The spatially flat FRW holographic model proposed in \cite{wd-cqg}
rests on two main assumptions: $(i)$ The dark energy density is
governed by the saturated holographic relationship, Eq.
(\ref{rhox}), with the infrared cutoff fixed by the Hubble radius,
i.e., $L = H^{-1}$. $(ii)$ Dark matter and dark energy do not
evolve independently of each other. They interact according to
\begin{equation}
\dot{\rho}_{M} + 3H \rho_{M} = Q \, \, , \quad \mbox{and} \quad
\dot{\rho}_{X} + 3H (1+w)\rho_{X} = - Q \, , \label{dotrhom}
\end{equation}
where $w = p_{X}/\rho_{X}$ stands for the equation of state
parameter of dark energy, and
\be Q =  \Gamma \, \rho_{X} \label{Qterm} \ee
is the interaction term where $\Gamma$ denotes the rate by which
$\rho_{X}$ changes as a result of the interaction. We assume
$\Gamma$ to be semipositive-definite. Note that if $Q$ were
negative, the transfer of energy would go from dark matter to dark
energy, in contradiction with the second law of thermodynamics
\cite{db-grg}. Further, use of the Layzer-Irvine equation on
nearly one hundred galaxy clusters strongly supports this view
\cite{abdalla-abramo}.

Interacting models were first proposed by Wetterich  to lower down
the value of the cosmological term \cite{wetterich}. Later on it
was proved efficient in  easing the cosmic coincidence problem
\cite{luca,ladw} and it was suggested that the interaction
(whatever form it might take) is not only likely but inevitable
\cite{brax,jerome}. The amount of literature on the subject is
steadily increasing -see, e.g., \cite{srd} and references therein.
Admittedly, the expression (\ref{Qterm}) is nothing but a useful
parametrization of the interaction. Given our poor understanding
of the nature of dark matter and dark energy, there is no clear
guidance to derive an expression for $Q$ from first principles.
This is why our approach will be purely phenomenological.

The model is fully specified by three quantities, e.g., the
current value of the Hubble rate, $H_{0}$, the dimensionless
density parameter $\Omega_{X} := 8\pi G \rho_{X}/(3H^{2})$ (or,
equivalently, $\Omega_{M}$), and $\Gamma$. Note that $c^{2}$ is
fixed by $c^{2} = \Omega_{X}$, as it can be readily checked.

The first assumption readily implies that $\Omega_{X}$ does not
vary with expansion, and that the ratio of energy densities, $r :=
\rho_{M}/\rho_{X}$, stays fixed in spatially flat FRW universes
($\Omega_{M} \, + \, \Omega_{X} = 1$) for any interaction. The
latter consequence greatly alleviates the coincidence problem
albeit, strictly speaking, it does not solve it in full because
the model cannot predict that $r \sim {\cal O}(1)$ (to the best of
our knowledge, no model is able to predict that). This feature of
$\Omega_{X}$ and $r$ being strictly constants may seem too strong;
however, one should bear in mind that both quantities would
slightly vary with the Universe expansion if the parameter $c^{2}$
in Eq. (\ref{rhox}) were allow to weakly depend on time, something
not at all unreasonable. Further, $r$ would not be constant if the
restriction to spatial flatness were relaxed. At any rate, we
shall take the conservative stance that both $c^{2}$ and $r$ do
not vary; thus, the number of free parameters of the model will be
kept to a minimum.

\begin{figure}[!htb]
  \begin{center}
    \begin{tabular}{ccc}
      \resizebox{90mm}{!}{\includegraphics{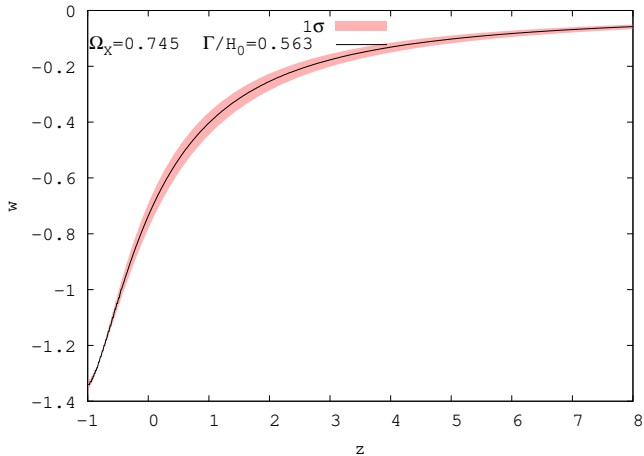}} \\
    \end{tabular}
    \caption{Evolution of the equation of state parameter of dark energy,
    Eq. (\ref{wde}), for the best fit model, up to $z = 8$. In this, as
    well as in subsequent figures, the red swath indicates the
    region obtained by including the $1\sigma$ uncertainties of
    the constrained parameters used in the calculation (in the present case,
    $\Omega_{X}$ and $\Gamma/H_{0}$).} \label{fig:w(z)}
  \end{center}
\end{figure}

\begin{figure}[!htb]
  \begin{center}
    \begin{tabular}{ccc}
      \resizebox{90mm}{!}{\includegraphics{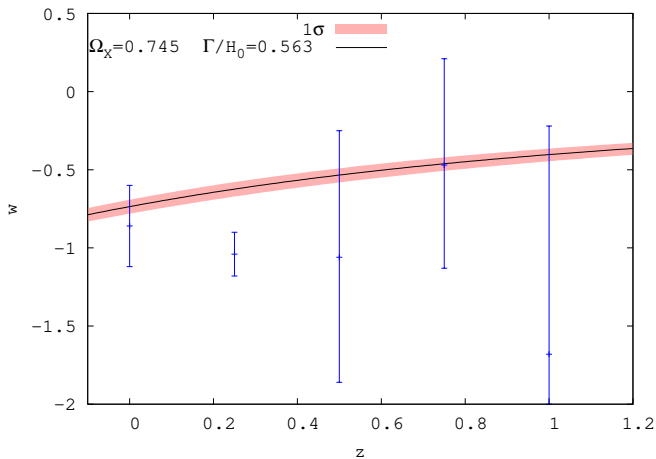}} \\
    \end{tabular}
    \caption{Evolution of the equation of state of dark energy for the best fit
    model up to $z = 1.2$. The observational data with their $2\sigma$ error bars are
    borrowed from \cite{Serra}. In plotting the curve no fit to these data was made.}
    \label{fig:w(z)_obs}
  \end{center}
\end{figure}

At first sight, the consequence of $\Omega_{X}$ being of order
unity also at early times might look worrisome. One may think that
a large dark energy component at that period would prevent the
formation of gravitationally bound objects. However, this is not
the case as
\be
w = - \frac{1\, +\, r}{r} \, \frac{\Gamma}{3H}
\label{wde}
\ee
is not constant, and for suitable choice of the ratio $\Gamma/H$
it tends to the equation of state of non-relativistic matter at
early times. Its evolution is governed by the Hubble rate which,
in the simplest case of $\Gamma$ being a constant, takes the form
\be H = H_{0}\left[\frac{\Gamma}{3 H_{0}r} + \left(1 -
\frac{\Gamma}{3 H_{0}r}\right)a^{-3/2}\right] \, ,
\label{hubble1}
\ee
which corresponds to a specific generalized Chaplygin gas
\cite{bento}. In last expression, the scale factor has been
normalized by setting $a_{0} = 1$.
Figures \ref{fig:w(z)} and \ref{fig:w(z)_obs} show the history of
the equation of state for the best fit values of the model up to
redshifts $8$ and $1.2$, respectively. Figure \ref{fig:w(z)}
illustrates that at high redshifts $w$ approaches zero
asymptotically. Figure \ref{fig:w(z)_obs} shows that, in
accordance with the analysis in \cite{Serra}, $w(z)$ varies little
at small redshifts.

\begin{figure}[!htb]
  \begin{center}
    \begin{tabular}{ccc}
      \resizebox{90mm}{!}{\includegraphics{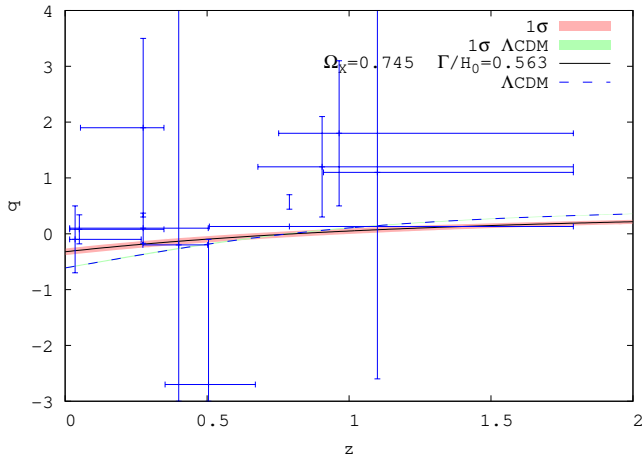}} \\
    \end{tabular}
    \caption{History of the deceleration parameter, according to
    Eq. (\ref{deceleration}) in terms of redshift for the best
    fit holographic model (solid line). The redshift at which the transition
    deceleration-acceleration occurs is approximately $0.80$.
    Also shown is the prediction of the $\Lambda$CDM model (dashed line).
    In this, as well as in subsequent figures, the green swath indicates the
    region obtained by including the $1\sigma$ uncertainties of
    the constrained parameters used in the calculation (in the
    present case just $\Omega_{M0}$). The data are borrowed from
    \cite{ruth}. In drawing the curves no fit to these data was made.}
    \label{fig:q(z)}
  \end{center}
\end{figure}

The deceleration parameter, $q := -\ddot{a}/(a \, H^{2})$, whose
evolution is illustrated in Fig. \ref{fig:q(z)}, obeys
\be q = \frac{1}{2}\left(1 \, -\, \frac{\Gamma}{Hr}\right) \, .
\label{deceleration} \ee
This expression implies that $q \rightarrow \textstyle{\frac{1}{2}}$
at high redshifts as it should, and that the transition from
deceleration to acceleration occurs at
\be
z_{tr} = \left( \frac{2 \Gamma}{3H_{0} \, r -
\Gamma}\right)^{2/3} \, - \, 1 \, ,
\label{trsnsition}
\ee
which yields $z_{tr} \simeq 0.80$ for the best fit values. It
should be noted that in \cite{dw-plb} the transition
deceleration-acceleration required that the $c^{2}$ varied, if
only very slowly. In the present case, the transition also occurs
for $c^{2} = $ constant (as, for simplicity, we are considering).
The difference stems from the fact that in  \cite{dw-plb} the
ratio $\Gamma/H$ was kept constant, while the present model has
$\Gamma =$ constant, instead.

\begin{figure}[!htb]
  \begin{center}
    \begin{tabular}{ccc}
      \resizebox{90mm}{!}{\includegraphics{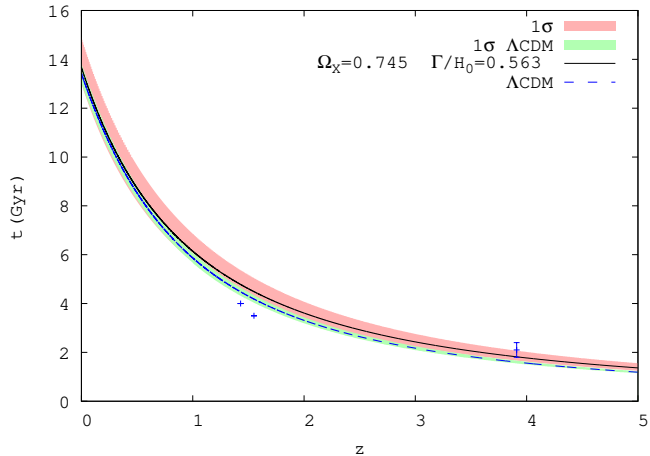}} \\
    \end{tabular}
    \caption{Dependence of the age of the Universe on redshift
    for the holographic model (solid line) and the $\Lambda$CDM model (dashed line).
    Also shown are the ages and redshifts of three old luminous objects,
    namely: galaxies LBDS 53W069, and LBDS 53091, and the
    quasar APM 08279+5255 - the latter with its $1\sigma$ error bar.
    In plotting the curves we have used the best fit value $H_{0} = 68.1 \pm 2.1 \, {\rm km/s/Mpc}$
    for the holographic model and $H_{0}=72.1^{+1.8}_{-1.9}\, {\rm km/s/Mpc}$ for the
    $\Lambda$CDM model.}
    \label{fig:t(z)}
  \end{center}
\end{figure}

The age of old luminous objects at high redshifts can constrain
cosmological models by simply requiring that their age at the
redshift they are observed do not exceed the age of the Universe
at that redshift. Figure \ref{fig:t(z)} depicts the dependence of
the age of the Universe on redshift for the best fit values of
both the holographic model and the $\Lambda$CDM model alongside
the age and redshift of three luminous old objects, namely:
galaxies LBDS 53W069 ($z = 1.43$, $t = 4.0$ Gyr)
\cite{dunlop_1999} and LBDS 53W091 ($z = 1.55$, $t = 3.5$ Gyr)
\cite{dunlop_1996,spinrad}, as well as the quasar APM 08279+5255
($z= 3.91$, $t= 2.1$ Gyr) \cite{hasinger,friaca}.
While the ages of the two first objects are lower than the ages of
the holographic model and the $\Lambda$CDM model at the
corresponding redshifts, the age of the quasar APM 08279+5255 lies
slightly further than $1\sigma$ beyond the age of the $\Lambda$CDM
model at $z= 3.91$. By contrast, the holographic model is
compatible at $1\sigma$ level with the age of the said quasar. The
tension between the APM quasar and the $\Lambda$CDM model has been
known for some time now (see \cite{friaca} and references therein)
and it has been revisited recently \cite{wei,cui}.
\\\\
\section {Observational constraints}
In this section we constrain the three free parameters
($\Omega_{X}$, $\Gamma/H_{0}$, and $H_{0}$) of the holographic
model presented above with observational data from SN Ia (557 data
points), the CMB-shift, BAO, and gas mass fractions in galaxy
clusters as inferred from x-ray data (42 data points), and the
Hubble rate (15 data points) to obtain the best fit values.  As
the likelihood function is defined by $ {\cal L} \propto \exp (-
\chi^{2}/2)$ the best fit follows from minimizing the sum $
\chi^{2}_{\rm total} = \chi^{2}_{sn} \, + \, \chi^{2}_{cmb}\, + \,
\chi^{2}_{bao} \, + \, \chi^{2}_{x-rays}+ \, \chi^{2}_{Hubble}$.
\\\\
\subsection {SN Ia}
We contrast the theoretical distance modulus
\be \mu_{th}(z_{i}) = 5\log_{10}\left(\frac{D_{L}}{{10{\rm
pc}}}\right)\, + \,\mu_{0}\, , \label{modulus} \ee
where $\mu_{0} = 42.38 \, -\, 5\log_{10} h$, with the observed
distance modulus $\mu_{obs}(z_{i})$ of the 557 supernovae type Ia
assembled in the Union2 compilation \cite{Amanullah}. The latter
data set is substantially richer than previous SN Ia compilations
and presents other advantages; mainly, the refitting of all light
curves with the SALT2 fitter and an upgraded control of systematic
errors. In (\ref{modulus}) $D_{L} = (1+z)
\int_{0}^{z}{\frac{dz'}{E(z';{\bf p})}}$ is the Hubble-free
luminosity distance, with ${\bf p}$ the model parameters
($\Omega_{X}$, $\Gamma/H_{0}$, and $H_{0}$), and $E(z; {\bf p}) :=
H(z; {\bf p})/H_{0}$.

The $\chi^{2}$ from the 557 SN Ia is given by
\be \chi^{2}_{sn}({\bf p}) = \sum_{i=1}^{557}
\frac{[\mu_{th}(z_{i}) \, - \,
\mu_{obs}(z_{i})]^{2}}{\sigma^{2}(z_{i})} \, , \label{chi2mu} \ee
where $\sigma_{i}$ stands for the 1$\sigma$ uncertainty associated
to the $i$th data point.

To eliminate the effect of the nuisance parameter $\mu_{0}$, which
is independent of the data points and the data set, we follow the
procedure of \cite{Nesseris-Perilovorapoulos_0} and obtain
$\tilde{\chi}^{2}_{sn} = \chi^{2\, (minimum)}_{sn} = 569.497$.
%
\subsection{CMB shift}
The CMB shift parameter measures the displacement of the first
acoustic peak of the CMB temperature spectrum with respect to the
position it would occupy if the Universe were accurately described
by the Einstein-de Sitter model. It is approximately
model-independent and given by \cite{wang-mukherjee,bond}
\begin{equation}\label{eq:CMBShift}
{\cal R} =\sqrt{\Omega_{M_{0}}}\int^{z_{rec}}_{0}\frac{dz}{E(z;
{\bf p)}} \, ,
\end{equation}
where $z_{rec}\simeq 1089$ is the redshift at the recombination
epoch. The 7-year WMAP data yields ${\cal R}(z_{rec})=1.725 \pm
0.018$ \cite{komatsu}. The best fit value of the model is ${\cal
R}(z_{rec})=1.753^{+0.033}_{-0.027}$. Minimization of
\be \chi^{2}_{cmb} ({\bf p})= \frac{({\cal R}_{th} \, - \, {\cal
R}_{obs})^{2}}{\sigma^{2}_{{\cal R}}} \, \label{chi2R} \ee
produces $\chi^{2\, (minimum)}_{CMB-shift} = 2.385$.
\\\\
\subsection{BAO}
Baryon acoustic oscillations can be traced to pressure waves at
the recombination epoch generated by cosmological perturbations in
the primeval baryon-photon plasma. They have been revealed by a
distinct peak in the large scale correlation function measured
from the luminous red galaxies sample of the Sloan Digital Sky
Survey (SDSS)  at $z = 0.35$ \cite{Eisenstein}, as well as in the
Two Degree Field Galaxy Redshift Survey (2dFGRS) at $z = 0.2$
\cite{Percival}. The peaks can be associated to expanding
spherical waves of baryonic perturbations. Each  peak introduces a
characteristic distance scale
\begin{equation}\label{eq:BAO}
D_{v}(z_{BAO})=\left[\frac{z_{BAO}}{H(z_{BAO})}
\left(\int^{z_{BAO}}_{0}\frac{dz}{H(z)}\right)^{2}\right]^{\frac{1}{3}}
\,
\end{equation}
(see Ref. \cite{Nesseris-Perilovorapoulos} for a pedagogical
derivation of this expression).

Data from SDSS  and 2dFGRS measurements  yield
$D_{v}(0.35)/D_{v}(0.2)=1.736 \pm 0.065 \, $ \cite{Percival}. The
best fit value for the holographic model is
$D_{v}(0.35)/D_{v}(0.2) = 1.642 \pm 0.003$, and minimization of
\be \chi^{2}_{bao} ({\bf p}) =
\frac{([D_{v}(0.35)/D_{v}(0.2)]_{th} \, - \,
[D_{v}(0.35)/D_{v}(0.2)]_{obs})^{2}}{\sigma^{2}_{D_{v}(0.35)/D_{v}(0.2)}}
\label{chi2BAO} \ee
gives $\chi^{2\, (minimum)}_{bao} = 2.089$.
\\\\
\subsection {Gas mass fraction}
Since the bulk of baryons in galaxy clusters are in the form of
hot x-ray emitting gas clouds (other baryon sources lagging far
behind in mass) the fraction of baryons in clusters, $f_{gas} : =
M_{gas}/M_{tot}$, results of prime interest for it seems a good
indicator of the overall cosmological ratio
$\Omega_{baryons}/\Omega_{M}$ and, up to a fair extent, it is
independent of redshift \cite{White}. The aforesaid fraction is
related to the cosmological parameters through  $f_{gas} \propto
d_{A}^{3/2}$, where \newline $d_{A}: =
(1+z)^{-1}\int_{0}^{z}{\frac{dz'}{H(z')}}$ stands for the angular
diameter distance to the cluster.

We used 42 Chandra measurements of  dynamically relaxed galaxy
clusters in the redshift interval $0.05 < z < 0.1$ \cite{Allen}.
To fit the data we have employed the empirical formula
\begin{equation}\label{eq:Allen}
f_{gas}(z)=\frac{K \, A \, \gamma\, b(z)}{1\, + \, s(z)}
\frac{\Omega_{B0}}{\Omega_{M0}}\left(\frac{d_{A}^{\Lambda
CDM}}{d_{A}}\right)^{3/2}
\end{equation}
(see Eq. (3) in Ref. \cite{Allen}) in which the $\Lambda$CDM model
is utilized as reference. Here, the parameters $K$, $A$, $\gamma$,
$b(z)$ and $s(z)$ model the amount of gas in the clusters. We fix
these parameters to their respective best fit values which can be
found in Ref. \cite{Allen}.

The $\chi^{2}$ function from the 42 galaxy clusters reads
\be
\chi^{2}_{x-rays}({\bf p}) = \sum_{i=1}^{42}
\frac{([f_{gas}(z_{i})]_{th} \, - \,
[f_{gas}(z_{i})]_{obs})^{2}}{\sigma^{2}(z_{i})} \, ,
\label{chi2fgas}
\ee
and its minimum value results to be $\chi^{2\, (minimum)}_{x-rays}
= 44.758$.

\begin{figure}[!htb]
  \begin{center}
    \begin{tabular}{ccc}
      \resizebox{90mm}{!}{\includegraphics{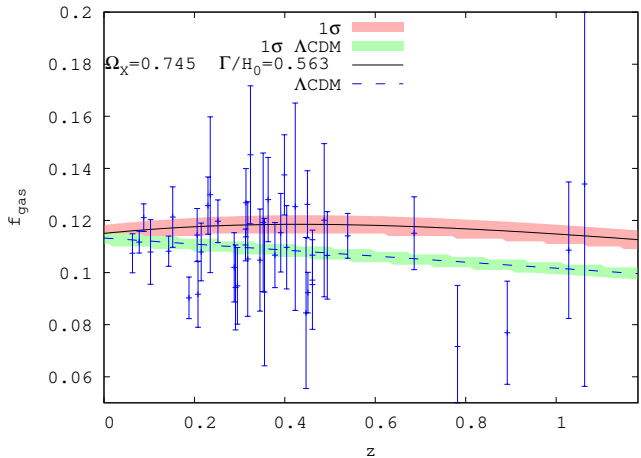}} \\
    \end{tabular}
    \caption{Gas mass fraction in 42 relaxed galaxy clusters vs.
    redshift. The solid and dashed curves correspond to the best fit
    models: holographic and $\Lambda$CDM, respectively.
    The data points with their error bars are taken from Table III
    in Ref. \cite{Allen}.}
    \label{fig:XRay}
  \end{center}
\end{figure}

Figure \ref{fig:XRay} shows the fit to the data.

\subsection {History of the Hubble parameter}
The history of the Hubble parameter, $H(z)$, is poorly constrained
though, recently, some high precision measurements by Riess {\em
et al.} at $z = 0$, obtained from the observation of 240 Cepheid
variables of rather similar periods and metallicities
\cite{Riess-2009}, and Gazta\~{n}aga {\em et al.}, at $z = 0.24,
\, 0.34, {\rm and}\,  0.43 \,$ \cite{gazta}, who used the BAO peak
position as a standard ruler in the radial direction, have
improved matters somewhat. To constrain the model we have employed
these four data alongside 11 less precise data, in the redshift
interval $0.1 \lesssim z \lesssim 1.8$, from Simon {\em et al.}
\cite{simon} and Stern {\em et al.} \cite{stern}, derived from the
differential ages of passive-evolving galaxies and archival data.

\begin{figure}[!htb]
  \begin{center}
    \begin{tabular}{ccc}
      \resizebox{90mm}{!}{\includegraphics{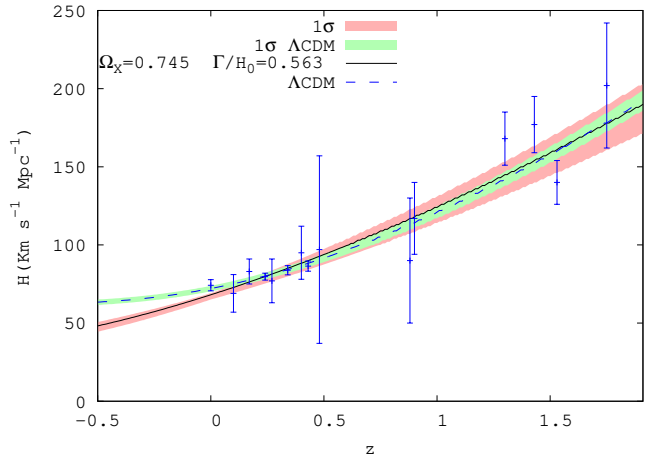}} \\
    \end{tabular}
    \caption{Plot of $H(z)$ for the best fit values of the holographic model
     (solid line) and the $\Lambda$CDM model (dashed line). The data
     points and error bars are borrowed from Refs. \cite{Riess-2009,gazta,simon}.}
    \label{fig:H(z)}
  \end{center}
\end{figure}

Minimization of
\be \chi^{2}_{Hubble}({\bf p}) = \sum_{i=1}^{15}
\frac{[H_{th}(z_{i}) \, - \,
H_{obs}(z_{i})]^{2}}{\sigma^{2}(z_{i})}
 \label{chi2hubble}
 \ee
provided us with $\chi^{2\, (minimum)}_{Hubble} = 11.897$ and
$H_{0} = 68.1 \pm 2.1 \,$km/s/Mpc as the best fit for the Hubble's
constant. Figure \ref{fig:H(z)} depicts the Hubble history
according to the best fit holographic model alongside the best
$\Lambda$CDM model.
\[\]

\begin{figure}[!htb]
  \begin{center}
    \begin{tabular}{cc}
      \resizebox{90mm}{!}{\includegraphics{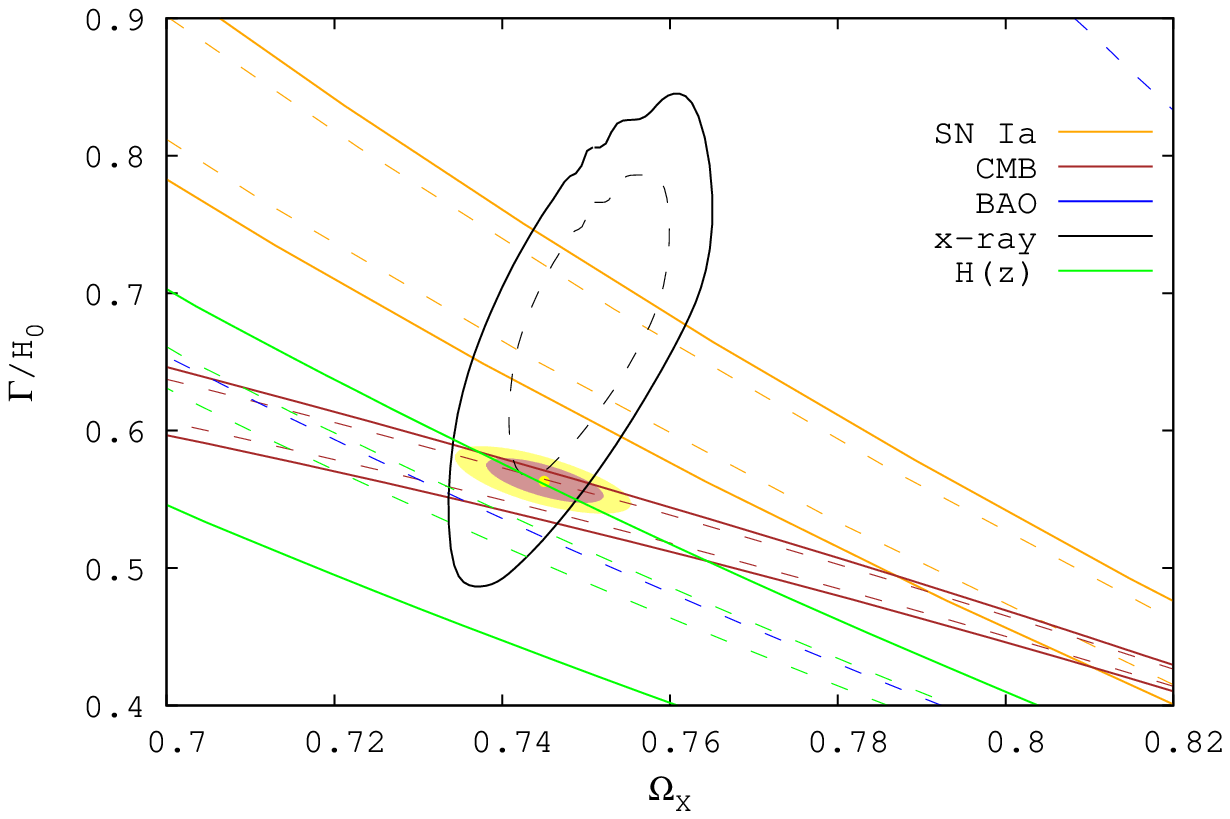}} &
      \resizebox{90mm}{!}{\includegraphics{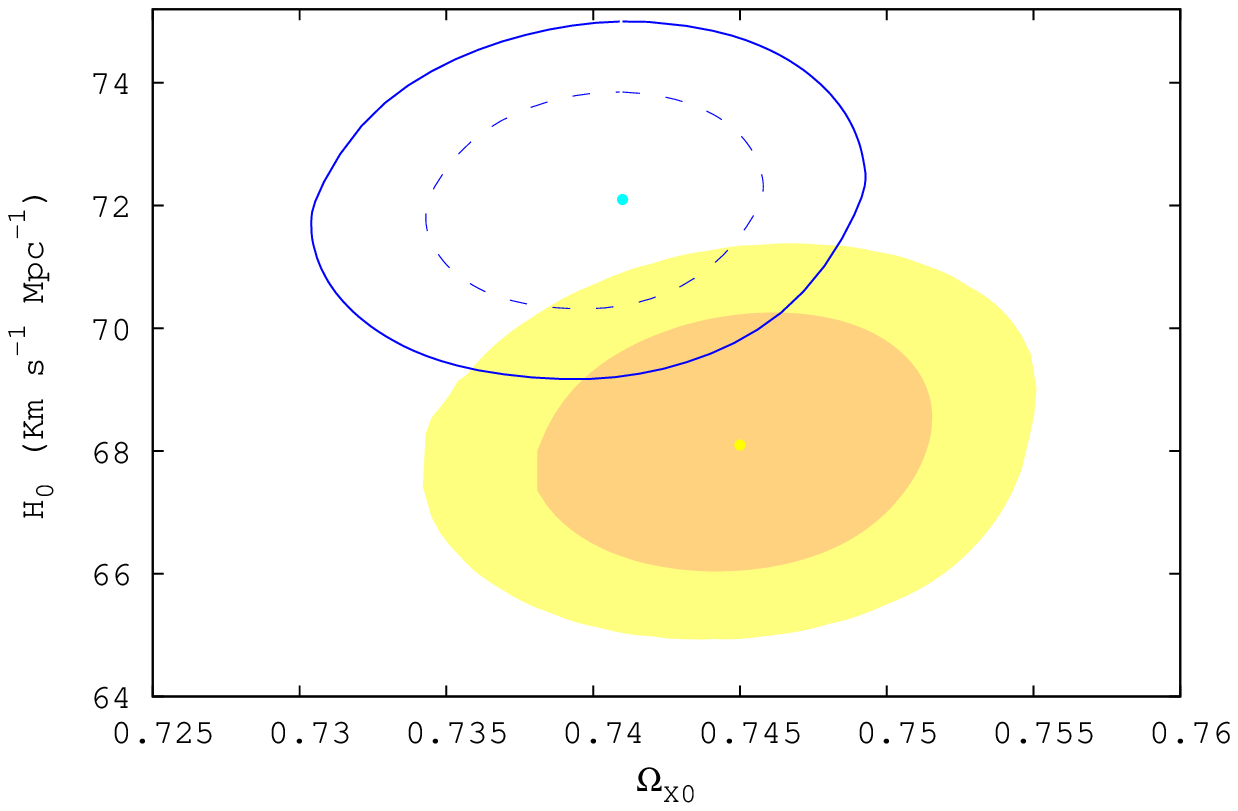}}\\
    \end{tabular}
    \caption{Left panel: the 68.3\% and 95.4\% confidence contours for the pair of free
     parameters  ($\Omega_{X}\,$, $\Gamma/H_{0})$ obtained by constraining the holographic
     model with SN Ia+CMB-shift+ BAO+x-ray+H(z) data. The joined constraints corresponding
     to $\chi^{2}_{total}$ are rendered as shaded contours. The no interacting case is
     largely disfavored by the data. Right panel: the 68.3\% and 95.4\% confidence contours
     for the pair ($\Omega_{X0}\,$, $H_{0}$) of the holographic model (shaded contours)
     and the $\Lambda$CDM model (blue contours). The solid points signal the location of
     the best fit values. Notice the overlap at $2\sigma$ confidence level between
     both models.}
    \label{fig:ellipses}
  \end{center}
\end{figure}

\begin{figure}[!htb]
  \begin{center}
    \begin{tabular}{ccc}
      \resizebox{58mm}{!}{\includegraphics{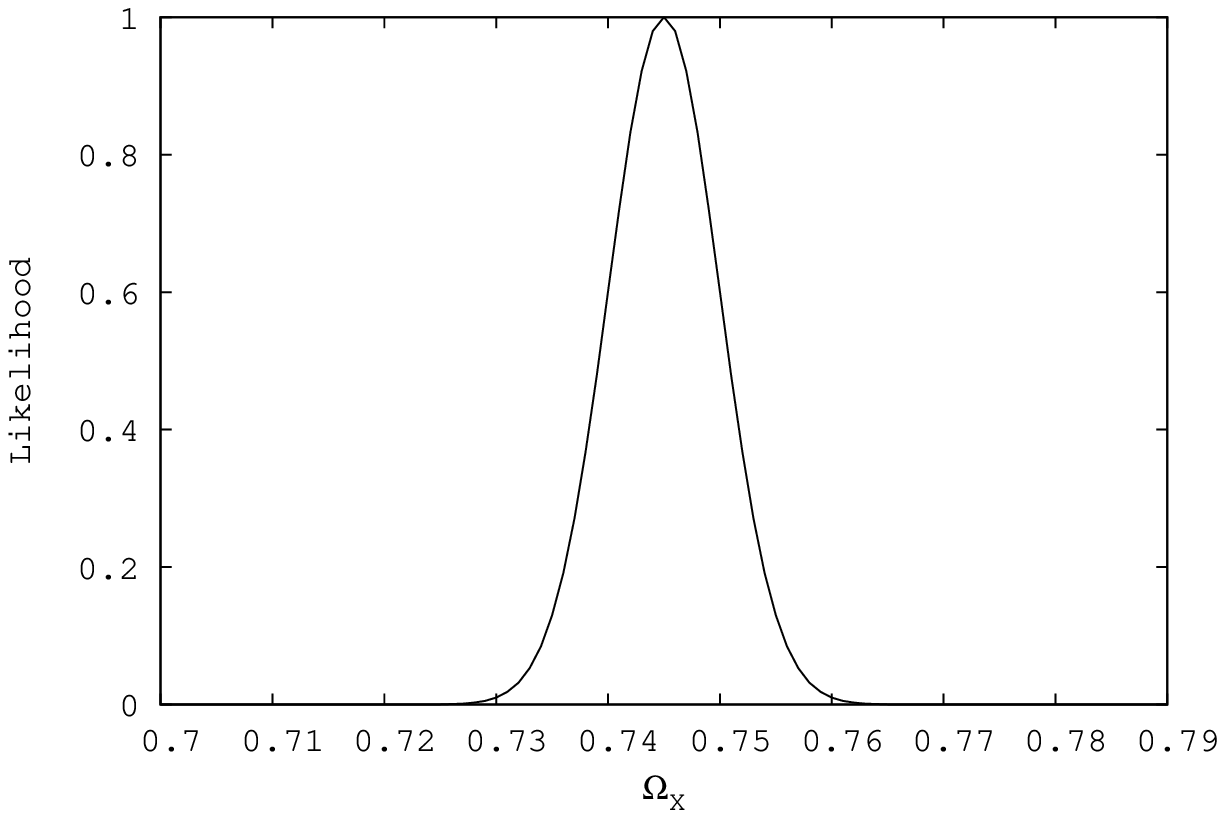}} &
      \resizebox{58mm}{!}{\includegraphics{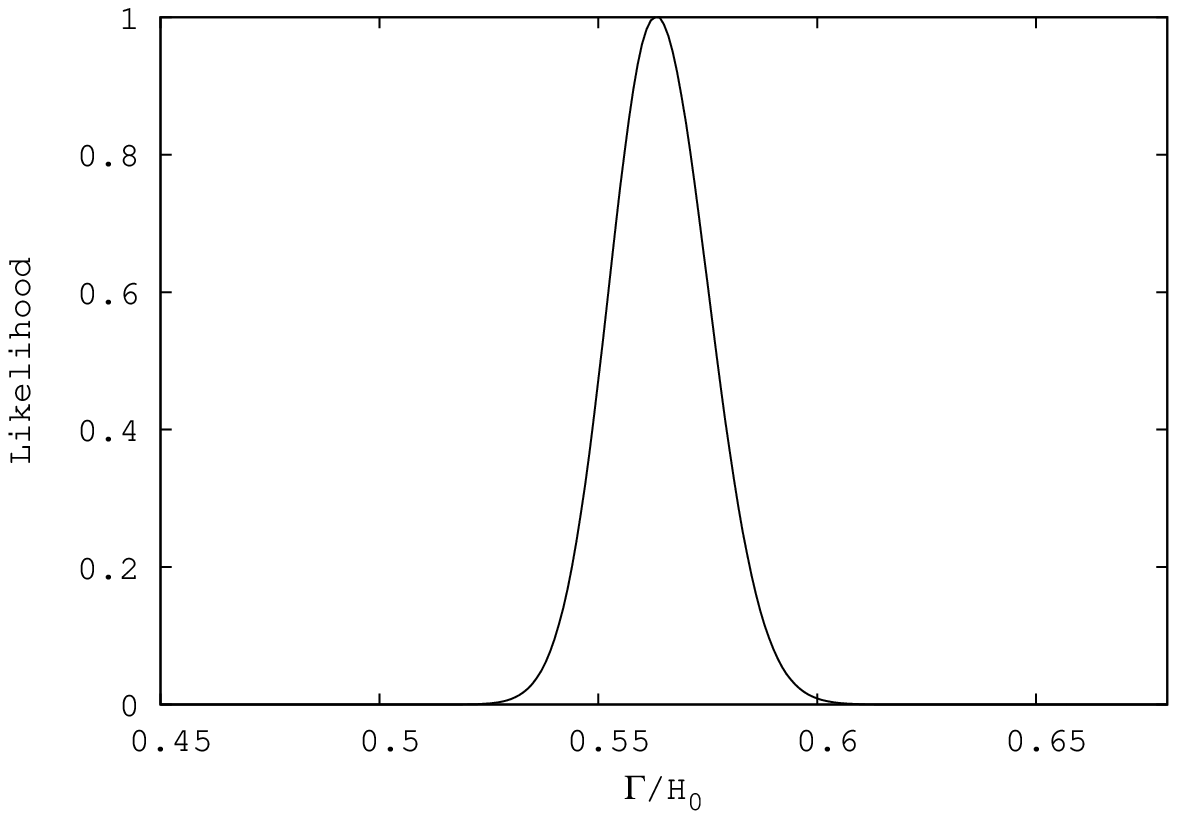}} &
      \resizebox{58mm}{!}{\includegraphics{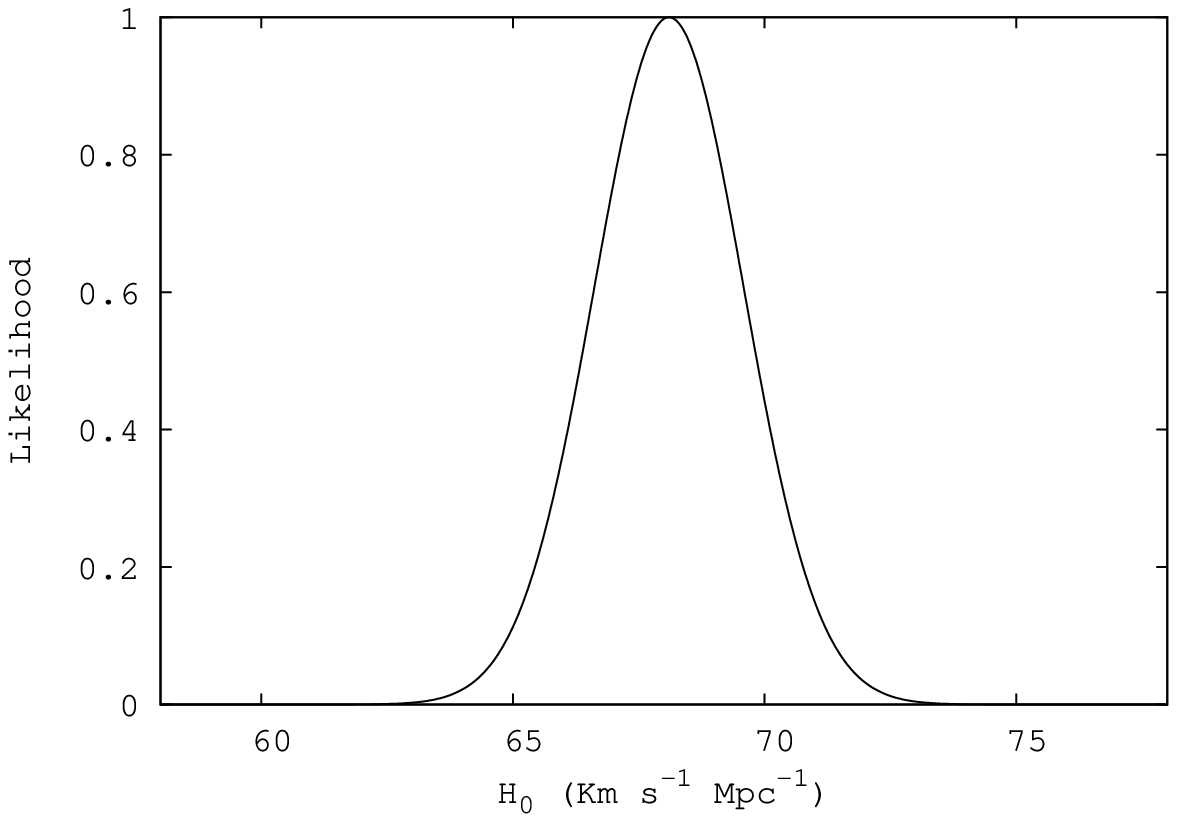}} \\
    \end{tabular}
    \caption{The normalized likelihoods of $\Omega_{X}$, $\, \Gamma/H_{0}$, and $H_{0}$.}
    \label{fig:likelihoods}
  \end{center}
\end{figure}

Figures \ref{fig:ellipses} and  \ref{fig:likelihoods} summarize
our analysis. The left panel of Fig. \ref{fig:ellipses} depicts
the 68.3\% and 95.4\% confidence contours for SN Ia (orange), CMB
shift (brown), BAO (blue), x-ray (black), and H(z) (green), in the
($\Omega_{X}, \, \Gamma/H_{0}$) plane. The joined constraints
corresponding to $\chi^{2}_{total}$ are shown as shaded contours.
The right panel depicts the  68.3\% and 95.4\% confidence regions
in the ($\Omega_{X0}, \, H_{0}$) plane of the holographic model
(shaded regions) and the $\Lambda$CDM model (blue contours). As it
is apparent, the models present a non-small overlap at $2\sigma$
level.

Figure \ref{fig:likelihoods} depicts the normalized likelihoods,
${\cal L} \propto \exp(-\chi_{total}^{2}/2)$, of the three free
parameters of the holographic model.

Altogether, by constraining the holographic model presented in
Section II with SN Ia, CMB-shif, BAO, x-rays, and H(z) data we
obtain $\Omega_{X}=0.745 \pm 0.007$, $\Gamma/H_{0}=0.563^{+
0.017}_{-0.015}$, and $H_{0} = 68.1 \pm 2.1\,$km/s/Mpc as best fit
parameters, with $\chi^{2}_{total}=630.627$. This value lies well
inside the $1\sigma$ interval ($\chi^{2}_{total}/dof\approx
1.03$). It should be noted that the no interacting case is
discarded at very high confidence level. This means no surprise at
all since for $\Gamma = 0$ the model reduces to the Einstein-de
Sitter ($\Omega_{M} =1$, $\Omega_{X} = 0$) and accordingly, as Eq.
(\ref{deceleration}) tells us, the transition from deceleration to
acceleration cannot occur.

Table \ref{table:chi2} shows the partial, total, and total
$\chi^{2}$ over the number of degrees of freedom of the
holographic model  along with the corresponding values for the
$\Lambda$CDM model. In the latter one has just two free
parameters, $\Omega_{M0}$ and $H_{0}$. Their best fit values after
constraining the model to the data are $\Omega_{M0} =
0.259^{+0.006}_{-0.005}$, and $H_{0} = 72.1^{+1.8}_{-1.9} \,
$km/s/Mpc, with $\chi^{2}_{total} = 593.142$.

\begin{table}
\begin{tabular}{ p{2.6 cm}| p{2.0 cm} p{1.6 cm} p{1.6 cm} p{1.8 cm}p{1.8 cm}p{1.8 cm} p{2.0 cm}}
\hline \hline
Model & $\chi^{2}_{sn}$ & $\chi^{2}_{cmb}$ & $\chi^{2}_{bao}$ &  $\chi^{2}_{x-rays}$ &  $\chi^{2}_{H}$ & $\chi^{2}_{{\rm total}}$ & $\chi^{2}_{{\rm total}}/dof$  \\
\hline
Holographic  & $569.497$ & $2.385$ & $2.089$ & $44.758$ & $11.897$ & $630.627$ & $\; \; 1.03$ \\
\hline
$\Lambda$CDM & $541.833$ & $0.013$ & $1.047$ & $41.527$ & $8.727$ & $593.142$ & $\; \; 0.97$ \\
\hline \hline
\end{tabular}
\normalsize
\medskip
\caption{$\chi^{2}$ values for the best fit holographic model
($\Omega_{X}=0.745 \pm 0.007$, $\; \Gamma/H_{0}=0.563^{+0.017}_{-
0.015}$, and $H_{0} = 68.1 \pm 2.1\,$km/s/Mpc), and the best fit
$\Lambda$CDM model ($ \Omega_{M0} = 0.259^{+0.006}_{-0.005}$, and
$H_{0}=72.1^{+1.8}_{-1.9}\,$ km/s/Mpc).} \label{table:chi2}
\end{table}

We see that the $\Lambda$CDM model fits the data better than the
holographic model in spite of having one parameter less. Thus, the
former model should be preferred on statistical grounds.
Nevertheless, this does not tell the whole story; the $\Lambda$CDM
cannot address the cosmic coincidence problem and has some tension
with the age of the old quasar APM 08279+5255. By contrast, the
holographic model answers the said problem and shows
compatibility, at $1\sigma$, with the age of the old quasar.

\section{Evolution of the growth function}
It is not unfrequent to find in the literature cosmological models
that differ greatly on their basic assumptions but, nevertheless,
present a rather similar dynamical behavior. It is, therefore,
rather hard to discriminate them at the background level. However,
their differences are more readily manifested at the perturbative
level (though, admittedly, the uncertainty in the corresponding
data are, in general, wider). This justifies our interest in
studying the evolution of the matter perturbations  of the
holographic model inside the horizon.

A prime tool in this connection is the growth function, defined as
\be f:= d \ln \delta_{M}/d \ln a \, , \label{gfunction} \ee
where $\delta_{M}$ denotes the density contrast of matter. In
order to derive an evolution equation for $f$, we start from the
energy balance for the matter component in the Newtonian
approximation
\begin{equation}
\dot{\delta}_{M} - \frac{k^{2}}{a^{2}}\, v_{M} = -
\frac{Q}{\rho_{M}}\, \delta_{M} + \frac{\hat{Q}}{\rho_{M}}\, .
\label{ebalm}
\end{equation}
Here,  $v_{M}$ is the velocity potential, defined by
$\hat{u}_{M\alpha} \equiv v_{M,\alpha}$, where  $u_{M\alpha}$ is
the matter four-velocity, and the hat means perturbation of the
corresponding quantity.

Recalling Eqs. (\ref{dotrhom}) and (\ref{Qterm}) and that $\Gamma$
and $r$ do not vary, we can write
\be
\dot{\delta}_{M} - \frac{k^{2}}{a^{2}}v_{M} = -
\frac{\Gamma}{r}\left(\delta_{M} - \delta_{X}\right) \, .
\label{ebalmg}
\ee
Usually, the density contrast of dark energy is neglected under
the assumption that dark energy does not cluster on small scales.
However, as forcefully argued by Park {\em et al.}
\cite{Park-Hwang}, the neglecting of $\delta_{X}$ can be fully
justified in the case of the cosmological constant only. At any
rate, in the present case the setting of $\delta_{X}$ to zero wold
be incorrect given the coupling between both energy components at
the background level (i.e., Eqs. (\ref{dotrhom})). It seems
therefore reasonable to include a coupling, at least
approximately, also at the perturbative level. The simplest
possibility is to assume a proportionality  $\delta_{X} = \alpha
\, \delta_{M}$ with a constant $\alpha$. As we shall see, the only
consistent choice for this constant (under the conditions that
$\Gamma$ and $r$ are held fixed) is $\alpha = 1$. Thus,
Eq.~(\ref{ebalmg}) becomes
\begin{equation}
\dot{\delta}_{M} - \frac{k^{2}}{a^{2}}v_{M} = -
\frac{\Gamma}{r}\left(1- \alpha\right)\delta_{M}\, .
\label{ebalmgal}
\end{equation}
An equation for $v_{M}$ follows from the momentum conservation of
the matter component. Assuming that there is no source term in the
matter rest frame, this equation takes the simple form
\begin{equation}
\dot{v}_{M} + \phi = 0\, , \label{mbalm}
\end{equation}
where $\phi$ is the Newtonian potential. Differentiation of
(\ref{ebalmgal}), use of (\ref{mbalm}) and (\ref{ebalmgal}), and
substitution of the scale factor for the time as independent
variable, leads to
\begin{equation}
\delta_{M}^{\prime\prime} + \frac{3}{2a}\left[1 +
\frac{\Gamma}{3Hr} +
\frac{2\left(1-\alpha\right)}{3}\frac{\Gamma}{Hr}\right]\delta_{M}^{\prime}
- \frac{3}{2a^{2}}\frac{r+\alpha}{r+1}\left[1 -
\frac{4\left(1-\alpha\right)}{3}
\frac{\Gamma}{Hr}\frac{r+1}{r+\alpha}\right]\delta_{M} = 0 \, ,
\label{dpp}
\end{equation}
where use of Friedmann's equation,  $4\pi G \rho_{m} =
\frac{3}{2}H^{2}\frac{r}{1+r}$, has been made; the prime means
derivative with respect to $a$.

For a vanishing $\Gamma$ we must recover the conventional
perturbation equation $\delta_{M}^{\prime\prime} +
\frac{3}{2a}\delta_{M}^{\prime} - \frac{3}{2a^{2}}\delta_{M} = 0$
with the growing solution $\delta_{M} \propto a$ for a dust
universe. Clearly, this is only feasible for $\alpha = 1$. With
this choice the fractional matter perturbation $\delta_{M}$
coincides with the total fractional energy density perturbation,
$\delta \equiv \frac{\hat{\rho}_{M} + \hat{
\rho}_{X}}{\rho_{M}+\rho_{X}}$. It follows that the basic matter
perturbation equation for the interacting holographic models
reduces to
\be \delta_{M}^{\prime\prime} + \frac{3}{2a}\left[1 +
\frac{\Gamma}{3Hr}\right]\delta_{M}^{\prime} -
\frac{3}{2a^{2}}\delta_{M} = 0\ . \label{dppfin} \ee
Replacing $\delta_{M}$ by the growth function $f$, last equation
becomes

\be f^{\prime}\,  + \, f^{2} + \, \frac{1}{2}\left(1 +
\frac{\Gamma}{Hr}\right)f - \frac{3}{2} = 0
 \,
\label{eqf}
\ee
with $f' : = df/d \ln a$. This has the advantage of being a first
order differential equation. Notice that in the absence of
interaction, $\Gamma = 0$, its solution is simply $f = 1$ as it
should, i.e., a dust dominated universe.

\begin{figure}[!htb]
  \begin{center}
    \begin{tabular}{ccc}
      \resizebox{120mm}{!}{\includegraphics{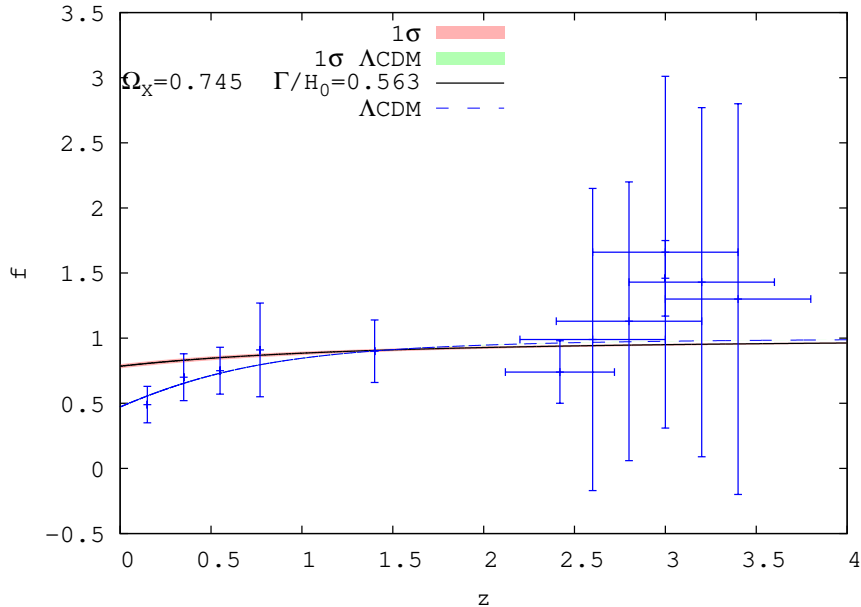}} \\
    \end{tabular}
    \caption{Growth function vs. redshift for the best fit holographic model
    (solid line). Also shown is the prediction of the $\Lambda$CDM model (dashed
    line). The observational data are borrowed from Table II in Ref. \cite{yungui}.
    In plotting the curves no fit to these data was made.}
    \label{fig:f(z)}
  \end{center}
\end{figure}

Figure \ref{fig:f(z)} depicts the evolution of the growth function
in terms of the redshift for the holographic as well as for the
$\Lambda$CDM model. The latter appears to fit the data below $z
\simeq 0.6$ better than the former. In particular, at $z = 0.15$
the best fit holographic model deviates $\Delta f = 0.3$
(corresponding to $3\sigma$) from the observed value (though it
falls within $1\sigma$ with the remaining data points) while the
best fit $\Lambda$CDM model falls within $1\sigma$ also at $z =
0.15$.

At any rate, it has been recently pointed out, from the
observation of nearby galaxies, that structure formation must have
proceed faster than predicted by the $\Lambda$CDM model
\cite{peebles-nature}. Clearly, slightly enhanced values of $f$ at
low  redshifts helps accelerate the formation of galaxies and
clusters thereof.

\section{Concluding remarks}
We constrained the interacting holographic model of Section II
with data from SN Ia, CMB shift, BAO, the gas mass fraction in
galaxy clusters, and $H(z)$. The parameters of the best fit model
are: $\Omega_{X}=0.745 \pm 0.007$, $\; \Gamma/H_{0}=0.563
^{+0.017}_{-0.015} \,$, and $H_{0} = 68.1 \pm 2.1\, $km/s/Mpc. We
have not included data of the growth function in the likelihood
analysis given the wide uncertainties of the current data.
However, we have derived the differential equation for $f$, Eq.
(\ref{eqf}), and integrated it numerically for the best fit model.

It conforms reasonably well to the observational data but not so
well as the $\Lambda$CDM model (best fit values: $\Omega_{M0} =
0.259^{+0.006}_{-0.005}$, $H_{0}=72.1^{+1.8}_{-1.9}\,$ km/s/Mpc)
does notwithstanding the latter has one less free parameter than
the former. However, the holographic model greatly alleviates the
cosmic coincidence problem and seems compatible at $1\sigma$ level
with the age of the old quasar APM 08279+5255. Besides, the
observational data from the CMB shift, BAO, x-ray, and some of the
$H(z)$ data, are not fully model independent owing to the fact
that  they are extracted with the help of the conventional
$\Lambda$CDM. This frequently makes  the latter tend to be
observationally favored over any other cosmological model.
Moreover, the BAO data are conventionally determined under the
assumption of purely adiabatic perturbations. However, as recently
argued \cite{mangilli}, should isocurvature components be present
the shape and location of the CMB acoustic peaks would be altered
and the data extracted from BAO affected.

Clearly, we must wait for more abundant, varied, and
model-independent accurate data to tell which of the two models
survives. If eventually neither of the two does, we should not be
so much disenchanted because, at any rate, this ``negative" result
would have narrowed significantly the parameter space of dark
energy.

\acknowledgments{We are indebted to Fernando Atrio-Barandela for
helpful comments on an earlier draft of this paper. ID research
was funded by the ``Universidad Aut\'{o}noma de Barcelona" through
a PIF fellowship. DP is grateful to the ``Departamento de
F\'{\i}sica de la Universidade Federal do Espirito Santo", where
part of this work was done, for financial support and warm
hospitality. This research was partly supported by the Spanish
Ministry of Science and Innovation under Grant
FIS2009-13370-C02-01, and the ``Direcci\'{o} de Recerca de la
Generalitat" under Grant 2009SGR-00164. Also, this work was
partially funded by CNPq (Brazil) and FAPES (Brazil).}

\end{document}